\documentstyle[twocolumn,aps]{revtex}

\input psfig.tex

\newcommand{\ltrsim}{\mathrel{\lower .3ex \rlap{$\sim$}
\raise .5ex\hbox{$<$}}}
\newcommand{\gttrsim}{\mathrel{\lower .3ex \rlap{$\sim$}
\raise .5ex\hbox{$>$}}}

\begin{document}
\draft

\twocolumn[\hsize\textwidth\columnwidth\hsize\csname %
@twocolumnfalse\endcsname

\title{
Violation of Luttinger's Theorem in the Two-Dimensional $t$-$J$ Model
}

\author{
W. O. Putikka$^{a,b*}$, M. U. Luchini$^c$ and R. R. P. Singh$^d$
}

\address{
$^a$Department of Physics, University of Cincinnati, Cincinnati, OH 45221-0011\\
$^b$Department of Physics, The Ohio State University, Mansfield, OH 44906\\
$^c$Department of Mathematics, Imperial College, London SW7 2BZ, United
Kingdom\\
$^d$Department of Physics, University of California, Davis, CA 95616\bigskip\\
}

\maketitle
\begin{abstract}
We have calculated the high temperature series for the momentum 
distribution function $n_{\bf k}$ of the 2D $t$-$J$ model to 
twelfth order in inverse temperature.  
By extrapolating the series to $T=0.2 J$ we searched for a Fermi
surface of the 2D $t$-$J$ model.  We find that three
criteria used for estimating the location of a Fermi
surface violate Luttinger's Theorem, implying the $t$-$J$ model does
not have an adiabatic connection to a non-interacting model.\\
\vspace{0.3in}
\end{abstract}
]

Models for two-dimensional strongly correlated electrons play a central role
in attempts to understand high temperature superconductors\cite{rice}.  
However, the
2D models themselves are at present poorly understood.  One of the main points
of interest in studies of 2D strongly correlated electrons is how similar the
2D models are to Fermi liquid theory, the standard model for conventional
metals\cite{pines}.  
Many-body calculations for conventional metals are generally
perturbative, assuming an adiabatic relation to a non-interacting model,
with low energy excitations describable by quasiparticles.

By summing a perturbative expansion to all orders Luttinger\cite{ward} 
was able to
show that a sharp Fermi surface can exist for interacting electrons.  He
defined the Fermi surface to be the locus of points in {\bf k}-space where the
renormalized single particle energy is equal to the zero temperature chemical
potential $E_{\bf k}=\mu$.  This requires the imaginary part of
the retarded self-energy to vanish on the Fermi surface.  Luttinger was able to show\cite{lutt}
${\rm Im}\Sigma_{\rm ret}(\omega)\propto(\omega-\mu)^2$, satisfying this requirement.
An immediate consequence of
this perturbative calculation is that the volumes (areas in 2D) enclosed by
the interacting and non-interacting Fermi surfaces are the same, a statement
generally known as Luttinger's Theorem.

Using high temperature series we investigated the momentum distribution function for the
2D $t$-$J$ model on a square lattice, 
with the Hamiltonian for the $t$-$J$ model given by
\begin{equation}
H=-tP\sum_{\langle ij\rangle,\sigma}\left(c_{i\sigma}^{\dagger}
c_{j\sigma} + c_{j\sigma}^{\dagger}c_{i\sigma}\right)P+J\sum_{
\langle ij\rangle}{\bf S}_i\cdot{\bf S}_j,
\end{equation}
where the sums are over pairs of nearest neighbor sites and the projection operators
$P$ eliminate from the Hilbert space states with doubly occupied sites.  We calculated
the high temperature series for the momentum distribution function to twelfth order in inverse
temperature $\beta=1/k_{B}T$, extending a previous eighth order calculation by Singh and Glenister\cite{singh}.
The definition of the
single spin momentum distribution function is
\begin{equation}
n_{\bf k}=\sum_{\bf r}n_{\bf r}{\rm e}^{i{\bf k}\cdot{\bf r}},
\end{equation}
with $n_{\bf r}=\langle c_{0\sigma}^{\dagger}c_{{\bf r}\sigma}\rangle$.
For the calculations reported here we fix $J/t=0.4$ and the electron density $n=0.8$.  A
wider range of parameters will be explored in a future publication.

To reach low temperatures we need to analytically continue the series
for $n_{\bf k}$.  A standard way to do this is to use Pad\'e
approximants.  However, for $n_{\bf k}$ the straightforward application of
Pad\'es does not work very well.  One way to improve the convergence of
Pad\'e approximants is to change the functional form before calculating Pad\'es\cite{gutt}.
Exactly which change to make
is difficult to know for unknown functions.  One way to proceed is
to use the function itself as a scaling function.  To do this we first form the high 
temperature series for
the ratio of two values of $n_{\bf k}$ with the two {\bf k}-points closely spaced and all
other parameters the same.  This is an exact calculation using the exact coefficients we
have for $n_{\bf k}$.  The series for the ratio is then analytically continued using
Pad\'e approximants down to $T=0.2J$ (for $J=1500$ K this is 300 K).  At each temperature 
all of the
ratios are referenced back to the zone center with an absolute scale for $n_{\bf k}$ set by
enforcing the sum rule $\sum_{\bf k}n_{\bf k}=n/2$.
Using this technique we collected data for $n_{\bf k}$ at 1326 {\bf k}-points in the irreducible
wedge of the square Brillouin zone for five temperatures $T/J=0.2$, $0.4$, $0.6$, $0.8$ and $1.0$.
The {\bf k}-points have
a uniform spacing of $\pi/50$ for $k_x$ and $k_y$. 
The accuracy of the $n_{\bf k}$ calculation varies with {\bf k}, but for $T=0.2J$ is approximately
$5\%$, for $T=0.4J$ approximately $3\%$ and for higher temperatures $1\%$ or less.  

In analyzing our results our main goals were to search for a possible Fermi surface 
of the 2D $t$-$J$
model and compare its area to the area of the tight-binding model Fermi 
surface to check Luttinger's Theorem.  Since we cannot
reach $T=0$, where the Fermi surface is defined by a sharp discontinuity in $n_{\bf k}$, we consider
three criteria which can be calculated for non-zero temperature and might be expected 
to smoothly
approach the $T=0$ Fermi surface as $T\rightarrow 0$.  The curves in 
{\bf k}-space we investigated are
defined by i) $n_{\bf k}=1/2$, ii) $dn_{\bf k}/dT=0$ and 
iii) $|{\bf \nabla}_{\bf k}n_{\bf k}|$ maximal.

A comparison of the locus in {\bf k}-space where $n_{\bf k}=1/2$ for the $t$-$J$ model at $T=0.2J$ to the tight-binding model
Fermi surface is shown in Fig. 1.  The area enclosed by the $t$-$J$ model curve is {\it smaller} than the area of the
tight-binding model Fermi surface.  This difference is not due to non-zero temperature.  Fig. 2 shows the
temperature dependence of the $t$-$J$ model $n_{\bf k}$ for ${\bf k}_F=(0.4266\pi, 0.4266\pi)$, the Fermi
momentum of the tight-binding model on the zone diagonal.  Clearly $n_{\bf k}$ for the $t$-$J$ model with
this momentum remains below $1/2$ for all temperatures.  The insert in Fig. 2 shows the
temperature dependence of the chemical potential $\mu(T)$ for the $t$-$J$ model 
with the same parameters as for $n_{\bf k}$.  The chemical potential shows little
variation for $T\ltrsim J$, appropriate for a degenerate Fermi system, with $\mu\approx 1.66t = 4.15J$.  

\begin{figure}[htb]
\centerline{\psfig{figure=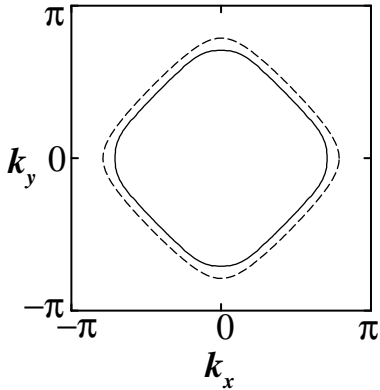,height=2in,angle=-90}}
\vspace{0.1in}
\caption{Comparison of the tight-binding Fermi surface (dashed line) 
to the locus of {\bf k}-points for the 2D $t$-$J$ model where $n_{\bf k}=1/2$ at $T=0.2J$ (solid line),
both at an electron density of $n=0.8$.  The area enclosed by $n_{\bf k}=1/2$ for the $t$-$J$ model
does not satisfy Luttinger's Theorem.}
\end{figure}

This comparison alone is not sufficient to claim a violation of Luttinger's
Theorem.  
In the absence of particle-hole symmetry $n_{\bf k}$ need not equal $1/2$ on
the Fermi surface\cite{randeria}.
The $n_{\bf k}=1/2$ criterion for the Fermi surface has
been widely applied in the past\cite{horsch} and is the 
simplest criterion to check before doing
more detailed calculations.  Our {\bf k}-resolution allows us to distinguish
$n_{\bf k}=1/2$ for the $t$-$J$ model from the Fermi surface of the tight-binding
model with approximately three data points separating the two curves.  Previous
calculations on small clusters\cite{horsch} 
did not have sufficient {\bf k}-resolution to make
this distinction.

An improved criterion for locating a Fermi surface has been proposed by Randeria {\it et al.}
\cite{randeria}.  
They proposed that the Fermi surface be identified with the locus of
{\bf k}-points where the temperature derivative of $n_{\bf k}$ is stationary,
$dn_{\bf k}/dT=0$.  
While the high temperature series for $dn_{\bf k}/dT$ can be
calculated directly from the series for $n_{\bf k}$, the resulting series is too
short to be extrapolated to low temperatures.  Alternatively, we consider the finite
difference approximation $\Delta n_{\bf k}/\Delta T$ where $\Delta T=0.2J$ and
$\Delta n_{\bf k}$ is found by directly subtracting the two $n_{\bf k}$'s for each
{\bf k}-point.  
Fig. 3 shows that this works very well for the tight-binding model
at $\bar{T}=0.3J$.  
From our data we find $\Delta n_{\bf k}/\Delta T$ centered on the
average temperatures $\bar{T}/J=0.3$, $0.5$, $0.7$ and $0.9$.  The results are shown
in Fig. 4.

\begin{figure}[htb]
\centerline{\psfig{figure=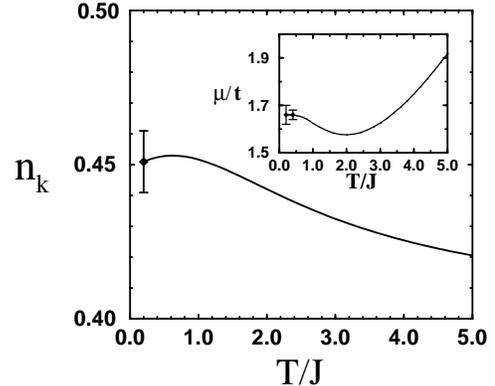,height=2in,angle=-90}}
\vspace{0.1in}
\caption{Temperature dependence of the $t$-$J$ model momentum distribution function $n_{\bf k}$ 
at the tight-binding Fermi wave vector 
${\bf k}_F=(0.4266\pi, 0.4266\pi)$ showing that on the tight-binding model Fermi
surface the $t$-$J$ model $n_{\bf k}<1/2$ for all temperatures.  Insert: Temperature dependence
of the $t$-$J$ model chemical potential.  For $T\gttrsim0.5J$ the error bars are the width of the line.}
\end{figure}

The results shown in Fig. 4 have three surprising features i) the area enclosed by
the curve where $\Delta n_{\bf k}/\Delta T=0$ is {\it larger} than the area enclosed by
the tight-binding Fermi surface, 
ii) {\bf k}-states with decreased occupancy as the temperature increases occur across
a broad region in the center of the Brillouin zone 
and iii) at low temperatures the distribution of {\bf k}-states 
with increased
occupancy as the temperature increases are strongly peaked on
the zone diagonal.  
Features ii) and iii) indicate the location of the low energy
excitations in the Brillouin zone.  Unlike the tight-binding model, where the low
energy excitations are confined to a narrow range of {\bf k}-points centered on the
Fermi surface, the low energy excitations for the $t$-$J$ model are spread throughout
the Brillouin zone.  The extremal values of $\Delta n_{\bf k}/\Delta T$ for the $t$-$J$
model are $\sim10$ times smaller than for the tight-binding model, implying a relatively
small density of states at any fixed momentum for the $t$-$J$ model.  
For $dn_{\bf k}/dT=0$ to give the correct Fermi surface 
particle-hole symmetry is required for low energies\cite{randeria}.
This may not be valid for the 2D $t$-$J$ model.  Thus the locus of {\bf k}-points
where $dn_{\bf k}/dT=0$ may not give the true Fermi surface.

The final criterion we considered for locating a Fermi surface for the
2D $t$-$J$ model is to follow the locus of {\bf k}-points where
$|{\bf\nabla}_{\bf k}n_{\bf k}|$ is maximal\cite{randeria,campuzano}.  This criterion only depends on
$n_{\bf k}$ having a sharp discontinuity at the Fermi surface for $T=0$.
For $T> 0$
the discontinuity will be smeared out, but we still expect $|{\bf\nabla}_{\bf k}n_{\bf k}|$
to be large near the Fermi surface since we have a degenerate Fermi
system.  As shown in Fig. 3 this criterion gives the proper Fermi surface
for the tight-binding model.  For the $t$-$J$ model we calculate 
$|{\bf\nabla}_{\bf k}n_{\bf k}|$ numerically from $n_{\bf k}$, with the results
shown in Fig. 5.

\begin{figure}[htb]
\centerline{\psfig{figure=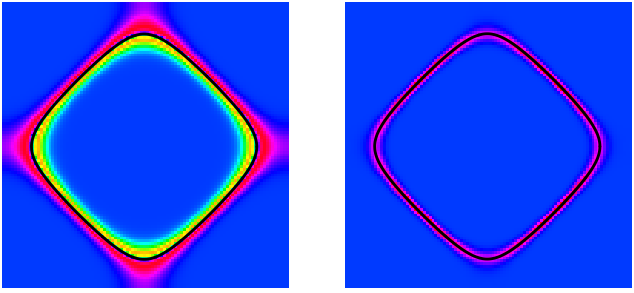,height=1.5in}}
\vspace{0.1in}
\caption{Full Brillouin zone plots for the tight-binding model.  Left: $\Delta n_{\bf k}/\Delta T$
for $\bar{T}=0.3J=0.12t$, Right: $|{\bf\nabla}_{\bf k}n_{\bf k}|$ for $T=0.2J=0.08t$. For both plots
blue is zero.  In the left plot the orange, yellow and green regions are negative with a minimum value of 
$\Delta n_{\bf k}/\Delta T=-0.9J^{-1}$ and the red and violet regions are positive with a maximum
value of $\Delta n_{\bf k}/\Delta T=0.6J^{-1}$.  In the right plot red and violet regions are positive
with a maximum value of $|{\bf\nabla}_{\bf k}n_{\bf k}|=6.8a$, where $a$ is the lattice spacing, here set 
to one.  Also, in both plots the solid black curve is the Fermi surface
of the tight-binding model, showing excellent agreement with both approximate techniques for locating
the Fermi surface.}
\end{figure}
\begin{figure}[htb]
\centerline{\psfig{figure=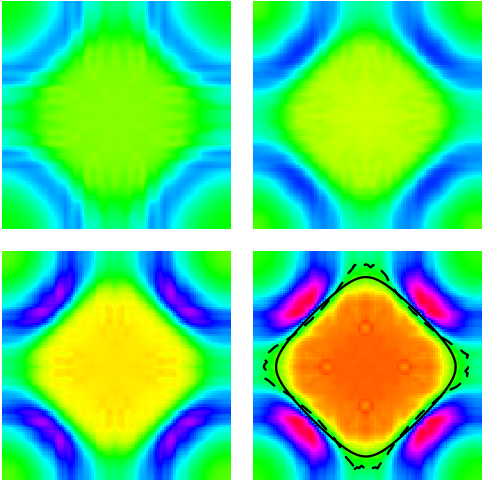,height=2.5in}}
\vspace{0.1in}
\caption{Full Brillouin zone plots of $\Delta n_{\bf k}/\Delta T$ for the $t$-$J$ model.
The color scale is the same for all four plots.  Orange, yellow and green are negative,
with a minimum value of $\Delta n_{\bf k}/\Delta T=-0.08J^{-1}$.  Blue, violet and red
are positive, with a maximum value of $\Delta n_{\bf k}/\Delta T=0.15J^{-1}$.  The plots
correspond to different temperatures.  Upper left: $\bar{T}=0.9J$, Upper right: $\bar{T}=0.7J$,
Lower left: $\bar{T}=0.5J$ and Lower right: $\bar{T}=0.3J$.  In the lower right plot the
solid black curve is the tight-binding model Fermi surface and the dashed curve is the
locus of {\bf k}-points where $\Delta n_{\bf k}/\Delta T=0$ for $\bar{T}=0.3J$.}
\end{figure}

At the lowest temperature shown in Fig. 5, $T=0.2J$, our results have two main
features i) the area enclosed by following a continuous locus of {\bf k}-points
along a ridge where $|{\bf\nabla}_{\bf k}n_{\bf k}|$ is maximal (as indicated by
the dotted line in Fig. 5) is {\it larger} than the area enclosed by the tight-binding
model Fermi surface and ii) $|{\bf\nabla}_{\bf k}n_{\bf k}|$ is strongly peaked
on the zone diagonal.  
However, compared to the tight-binding model, 
the maximum value
of $|{\bf\nabla}_{\bf k}n_{\bf k}|$ for the $t$-$J$ model at
$T=0.2J$ is $\sim10$ times smaller.
The shape of the area enclosed by the dotted line in Fig. 5
is similar to that for the tight-binding model with a next-nearest neighbor hopping
term, but this modification to the tight-binding Hamiltonian can only change the
shape of the Fermi surface, not its area.  Thus our results for the $t$-$J$ model
cannot be modeled as a band of non-interacting electrons with a modified
band structure.  

\begin{figure}[htb]
\centerline{\psfig{figure=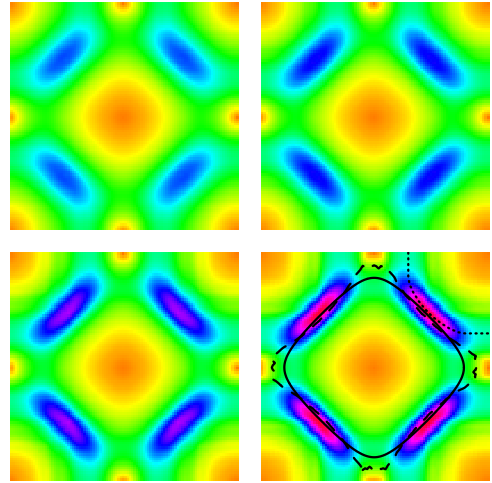,height=2.5in}}
\vspace{0.1in}
\caption{Full Brillouin zone plots of $|{\bf\nabla}_{\bf k}n_{\bf k}|$ for the $t$-$J$ model.
All of the colors are values of $|{\bf\nabla}_{\bf k}n_{\bf k}|\ge0$, with orange in the zone
center, around ${\bf k} = (\pi, \pi)$ and around ${\bf k} = (\pi, 0)$ being zero.  Red is
the maximum value $|{\bf\nabla}_{\bf k}n_{\bf k}|=0.43a$, where $a$ is the lattice spacing,
here set to one.  The plots are for different temperatures.  Upper left: $T=0.8J$, Upper right:
$T=0.6J$, Lower left: $T=0.4J$ and Lower right: $T=0.2J$.  In the lower right plot the
solid line is the tight-binding Fermi surface, the dashed line is the locus of {\bf k}-points from Fig. 4
where $\Delta n_{\bf k}/\Delta T=0$ and the dotted line is the locus of {\bf k}-points where 
$|{\bf\nabla}_{\bf k}n_{\bf k}|$ is maximal (only one of four branches is shown).}
\end{figure}

The main result of our calculation is that the 2D $t$-$J$ model violates Luttinger's
Theorem.  This means the ground state of the 2D $t$-$J$ model is not adiabatically
related to a non-interacting model.
Also, the distribution of low energy
excitations revealed by $\Delta n_{\bf k}/\Delta T$ suggests that quasiparticles
cannot describe all of the low energy degrees of freedom of the 2D $t$-$J$ model.
However, our results are not sufficient to uniquely determine the ground state
of the 2D $t$-$J$ model.

The simplest way to not have an adiabatic connection to
non-interacting electrons is for the 2D $t$-$J$ model to have an ordered ground state
with a different symmetry than non-interacting electrons.  The low temperature
growth of peaks in $\Delta n_{\bf k}/\Delta T$ and $|{\bf\nabla}_{\bf k}n_{\bf k}|$
suggests some kind of order is developing in the $t$-$J$ model.  The entropy of the
$t$-$J$ model\cite{ent} 
also starts to decrease at $T\sim J$.
At present it is not possible to determine the precise nature of this order.  We know from the spin
correlation function\cite{sq} 
and the antiferromagnetic correlation length\cite{spincharge} 
that we do not
have long range spin order.  Concomitantly, we do not observe 
hole pockets in $n_{\bf k}$\cite{holepock}.
For the $t$-$J$ model the charge fluctuations are suppressed, with no indication of
short wavelength structure in the charge density\cite{nq}.  
The location of peaks in
$\Delta n_{\bf k}/\Delta T$ and $|{\bf\nabla}_{\bf k}n_{\bf k}|$ along the zone diagonal
is consistent with the location of gap nodes for $d_{x^2-y^2}$ superconducting 
fluctuations.  If this is correct, superconducting fluctuations at low temperatures
are developing in the 2D $t$-$J$ model from a higher temperature state which cannot
be described as a Fermi liquid.  For $T\sim J$ the $t$-$J$ $n_{\bf k}$ can be modeled
by assuming spin-charge separation\cite{nq,nkfit}. 
However, there is no clear reason for peaks to appear in 
$\Delta n_{\bf k}/\Delta T$ or $|{\bf\nabla}_{\bf k}n_{\bf k}|$ for a spin-charge
separated state.

Other theoretical approaches have found results similar
to our data.  In particular $SU(2)$ gauge theory 
calculations\cite{su2} and phenomenological
models of preformed pairs\cite{preform} 
or of a d-wave ground state\cite{dwave} find Fermi arcs centered
on the zone diagonal and the Fermi surface gapped or destroyed by strong scattering near
${\bf k}=(\pi,0)$.  These theories are motivated by the pseudogap observed in ARPES\cite{arpes}.  
The largest value of $|{\bf\nabla}_{\bf k}n_{\bf k}|$ and the sharpest
region for $\Delta n_{\bf k}/\Delta T=0$ form arcs centered on the
zone diagonal with the same outward curvature as the ARPES data.  However, the
maximum value of 
$|{\bf\nabla}_{\bf k}n_{\bf k}|$ is substatially smaller than what
might be expected for a true discontinuity unless there is an energy
scale in the $t$-$J$ model smaller than $0.2J=300$K.
Our data also are similar to the ``cold spots'' proposed by 
Ioffe and Millis\cite{coldspot} and
quasiparticle decay as discussed by Laughlin\cite{qpdecay}, 
plus calculations for
$t$-$J$ ladders that show gapped and non-gapped features 
in $n_{\bf k}$\cite{ladder}.

ARPES experiments do not directly measure $n_{\bf k}$, though attempts have been made to
integrate the ARPES data over frequency, giving a value proportional to
$n_{\bf k}$, but with an unknown scale factor\cite{randeria}.  We can make a qualitative
comparison of our results to features of the ARPES data\cite{arpesrev}
i) $|{\bf\nabla}_{\bf k}n_{\bf k}|$ maximal gives a locus in {\bf k}-space 
with the same shape as the
Fermi surface found in ARPES experiments, ii) without Luttinger's Theorem, the peaks observed
by ARPES in the normal state\cite{randeria}
do not have to be quasiparticles and iii) we see a large region
with low energy excitations, similar to the flat background extending up to the chemical 
potential in ARPES data\cite{randeria,campuzano,arpes,arpesrev}.

In conclusion, we find that the $t$-$J$ model violates Luttinger's Theorem.  Using 
$|{\bf\nabla}_{\bf k}n_{\bf k}|$ as the least biased way to search for a Fermi surface we find 
the area enclosed by the curve where $|{\bf\nabla}_{\bf k}n_{\bf k}|$ is maximal to be larger 
than the area enclosed by the Fermi surface of the tight-binding model.  We also found structure
in the low energy excitations of the 2D $t$-$J$ model which might be due to $d_{x^2-y^2}$
superconducting fluctuations.

\noindent
{\it Acknowledgements.}
This work was partially supported by NSF grants DMR-9357199 (WOP) and
DMR-9616574 (RRPS). WOP thanks the ETH-Z\"urich for hospitality while
part of this work was being completed.\bigskip

\noindent
$^*$Current address.

\end{document}